\documentclass[]{llncs}

\usepackage[T1]{fontenc}
\usepackage[english]{babel}
\usepackage{amssymb,amsmath}
\usepackage{stmaryrd}
\usepackage{graphicx}
\usepackage{url}

\newcommand{\NN}{\mathbb{N}}
\newcommand{\ZZ}{\mathbb{Z}}
\newcommand{\ACA}{\mathcal{A}} 
\newcommand{\ACN}{\mathcal{N}} 
\newcommand{\ACQ}{\mathcal{Q}} 
\newcommand{\ACC}{\mathfrak{C}} 
\newcommand{\Conf}{\operatorname{Conf}} 

\begin{document}
\title{L-Convex Polyominoes are Recognizable in Real Time by 2D Cellular Automata}
\author{Ana\"el Grandjean \and Victor Poupet}
\authorrunning{A.~Grandjean and V.~Poupet}
\institute{LIRMM, Universit\'e Montpellier 2\\
161 rue Ada, 34392 Montpellier, France\\
\url{victor.poupet@lirmm.fr}, \url{anael.grandjean@lirmm.fr}
}
\maketitle

\begin{abstract}
	A polyomino is said to be L-convex if any two of its cells are connected by a 4-connected inner path that changes direction at most once. The 2-dimensional language representing such polyominoes has been recently proved to be recognizable by tiling systems by S.~Brocchi, A.~Frosini, R.~Pinzani and S.~Rinaldi. In an attempt to compare recognition power of tiling systems and cellular automata, we have proved that this language can be recognized by 2-dimensional cellular automata working on the von Neumann neighborhood in real time.
	
	Although the construction uses a characterization of L-convex polyominoes that is similar to the one used for tiling systems, the real time constraint which has no equivalent in terms of tilings requires the use of techniques that are specific to cellular automata.
\end{abstract}

\section*{Introduction}

Two-dimensional cellular automata and tiling systems are two different models that can be considered to recognize classes of two-dimensional languages (or picture languages). Although they share some similarities such as locality and uniformity, the two models are fundamentally different.

Tiling systems as language recognizers were introduced by D.~Giammarresi and A.~Restivo in 1992 \cite{giammarresi92} and are based on the model of tile sets introduced by H.~Wang \cite{wang61}. The strength of the model lies in its inherent non-determinism. The system itself is a set of local rules describing valid image patterns and a picture language is recognized by the system if it is the image by a projection of the set of configurations that verify all local rules.

Cellular automata on the contrary are deterministic dynamical models. Introduced in the 1940s by S.~Ulam and J.~von Neumann \cite{neumann66} to study self replication in complex systems they were rapidly considered as computation models and language recognizers \cite{smithIII72}. Contrary to some other classical computation models that inherently work on words, they can be considered naturally in any dimension (the original cellular automata studied by Ulam and von Neumann were 2-dimensional) and are therefore particularly well suited to picture languages. Language recognition is performed by encoding the input in an initial configuration and studying the (deterministic) evolution of the automaton from that configuration. Time and space complexities can be defined in the usual way.

Because tiling systems lack dynamic behavior, some picture languages that can be recognized by cellular automata with minimal space and time complexity (in real time) cannot be recognized by tiling systems, such as the language of square pictures with vertical symmetry.

Conversely, the non-determinism of tiling systems should allow the recognition of languages that cannot be recognized by cellular automata in low time complexities. It is straightforward for instance to verify that the language considered in \cite{terrier99} as an example of language that cannot be recognized in real time by a cellular automaton working on the Moore neighborhood but can be recognized on the von Neumann neighborhood can be recognized by a tiling system, thus proving that tiling systems and real time cellular automata on the Moore neighborhood are incomparable.

Because the language of L-convex polyominoes was recently proved to be recognizable by tiling systems when it was previously though not to be, we decided to investigate its recognizability by real time von Neumann neighborhood cellular automata. Although the language was also recognized by cellular automata, the construction turned out to be quite different from the case of tiling systems and used some techniques specific to cellular automata (and possibly von Neumann neighborhood cellular automata). This article describes said construction.

\section{Definitions}

\subsection{Cellular Automata}

\begin{definition}[Cellular Automaton]
	A \emph{cellular automaton} (CA) is a quadruple $\ACA = (d, \ACQ, \ACN, \delta)$ where
	\begin{itemize}
		\item $d\in\NN$ is the dimension of the automaton~;
		\item $\ACQ$ is a finite set whose elements are called \emph{states}~;
		\item $\ACN$ is a finite subset of $\ZZ^d$ called \emph{neighborhood} of the automaton~;
		\item $\delta: \ACQ^{\ACN} \rightarrow \ACQ$ is the \emph{local transition function} of the automaton.
	\end{itemize}
\end{definition}

\begin{definition}[Configuration]
	A \emph{$d$-dimensional configuration} $\ACC$ over the set of states $\ACQ$ is a mapping from $\ZZ^d$ to $\ACQ$.
	
	The elements of $\ZZ^d$ will be referred to as \emph{cells} and the set of all $d$-dimensional configurations over $\ACQ$ will be denoted as $\Conf_d(\ACQ)$.
\end{definition}

Given a CA $\ACA = (d, \ACQ, \ACN, \delta)$, a configuration $\ACC\in\Conf_d(Q)$ and a cell $c\in \ZZ^d$, we denote by $\ACN_\ACC(c)$ the neighborhood of $c$ in $\ACC$~:
\begin{displaymath}
	\ACN_\ACC(c) : \left\{\begin{array}{rcl}
		\ACN & \rightarrow & \ACQ \\
		n & \mapsto & \ACC(c+n)
	\end{array}\right.
\end{displaymath}

From the local transition function $\delta$ of a CA $\ACA = (d, \ACQ, \ACN, \delta)$, we can define the \emph{global transition function of the automaton} $\Delta: \Conf_d(\ACQ) \rightarrow \Conf_d(\ACQ)$ obtained by applying the local rule on all cells~:
\begin{displaymath}
	\Delta(\ACC) = \left\{\begin{array}{rcl}
		\ZZ^d & \rightarrow & \ACQ \\
		c & \mapsto & \delta(\ACN_\ACC(c))
	\end{array}\right.
\end{displaymath}
The action of the global transition rule makes $\ACA$ a dynamical system over the set $\Conf_d(\ACQ)$. Because of this dynamic, in the following we will identify the CA $\ACA$ with its global rule so that $\ACA(\ACC)$ is the image of a configuration $\ACC$ by the action of the CA $\ACA$, and more generally $\ACA^t(\ACC)$ is the configuration resulting from applying $t$ times the global rule of the automaton from the initial configuration $\ACC$.

\begin{definition}[Von Neumann and Moore Neighborhoods]
	In $d$ dimensions, the most commonly considered neighborhoods are the von Neumann neighborhood $\ACN_{\operatorname{vN}} = \{c \in \ZZ^d,\ ||c||_1 \leq 1\}$ and the Moore neighborhood $\ACN_{\operatorname{M}} = \{c \in \ZZ^d, ||c||_\infty \leq 1\}$. Figure~\ref{fig:neighborhoods} illustrates these two neighborhoods in 2 dimensions.

\begin{figure}
	\centering
	\begin{minipage}{0.35\textwidth}
		\centering
		\includegraphics[scale=1]{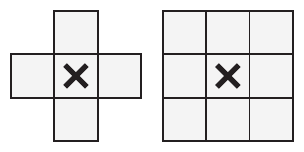}
		\caption{The von Neumann (left) and Moore (right) neighborhoods in 2 dimensions.}
		\label{fig:neighborhoods}
	\end{minipage}\hfill
	\begin{minipage}{0.60\textwidth}
		\centering
		\includegraphics[scale=1]{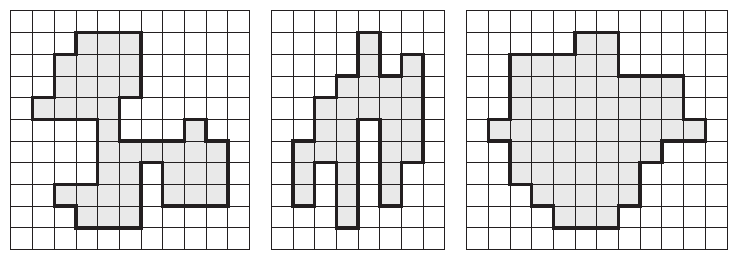}
		\caption{Three polyominoes. The center and right ones are vertically convex, the right one is HV-convex.}
		\label{fig:polyominoes}
	\end{minipage}
\end{figure}

\end{definition}

\subsection{Picture Recognition}

From now on we will only consider 2-dimensional cellular automata (2DCA),  and the set of cells will always be $\ZZ^2$.

\begin{definition}[Picture]
	For $n, m\in \NN$ and $\Sigma$ a finite alphabet, an \emph{$(n, m)$-picture} (picture of width $n$ and height $m$) over $\Sigma$ is a mapping
	\begin{displaymath}
		p: \llbracket 0, n-1\rrbracket \times \llbracket 0, m-1\rrbracket \rightarrow \Sigma
	\end{displaymath}
	
	$\Sigma^{n, m}$ denotes the set of all $(n, m)$-pictures over $\Sigma$ and $\Sigma^{*,*} = \bigcup_{n, m\in \NN} \Sigma^{n, m}$ the set of all pictures over $\Sigma$. A \emph{picture language} over $\Sigma$ is a set of pictures over $\Sigma$.
\end{definition}

\begin{definition}[Picture Configuration]
	Given an $(n, m)$-picture $p$ over $\Sigma$, we define the \emph{picture configuration} associated to $p$ with quiescent state $q_0\notin \Sigma$ as
	\begin{displaymath}
		\ACC_{p, q_0}: \left\{\begin{array}{rcl}
			\ZZ^2 & \rightarrow & \Sigma \cup \{q_0\} \\
			x, y & \mapsto & \left\{\begin{array}{rl}
				p(x, y) & \qquad\textrm{if $(x, y)\in \llbracket 0, n-1\rrbracket \times \llbracket 0, m-1\rrbracket$} \\
				q_0 & \qquad\textrm{otherwise}
			\end{array}\right.
		\end{array}\right.
	\end{displaymath}
\end{definition}

\begin{definition}[Picture Recognizer]
	Given a picture language $L$ over an alphabet $\Sigma$, we say that a 2DCA $\ACA=(2, \ACQ, \ACN, \delta)$ such that $\Sigma \subseteq \ACQ$ recognizes $L$ with quiescent state $q_0\in \ACQ\setminus \Sigma$ and accepting states $\ACQ_a \subseteq \ACQ$ in time $\tau: \NN^2 \rightarrow \NN$ if, for any picture $p$ (of size $n\times m$), starting from the picture configuration $\ACC_{p,q_0}$ at time 0, the origin cell of the automaton is in an accepting state at time $\tau(n, m)$ if and only if $p\in L$. Formally,
	\begin{displaymath}
		\forall n, m\in \NN, \forall p \in \Sigma^{n, m},\quad
		\ACA^{\tau(n, m)}(\ACC_{p, q_0})(0,0) \in \ACQ_a \Leftrightarrow p \in L
	\end{displaymath}
\end{definition}

Because cellular automata work with a finite neighborhood, the state of the origin cell at time $t$ (after $t$ actions of the global rule) only depends on the initial states on the cells in $\ACN^t$, where $\ACN^0 = \{0\}$ and for all $n$, $\ACN^{n+1} = \{x + y,\ x \in \ACN^n, y\in \ACN\}$. The real time function is informally defined as the smallest time such that the state of the origin may depend on all letters of the input~:

\begin{definition}[Real Time]
	Given a neighborhood $\ACN\subset \ZZ^d$ in $d$ dimensions, the real time function $\tau_\ACN: \NN^d \rightarrow \NN$ associated to $\ACN$ is defined as
\begin{displaymath}
	\tau_\ACN(n_1, n_2, \ldots, n_d) = \min\{t, \llbracket 0, n_1-1\rrbracket \times \llbracket 0, n_2-1\rrbracket \times\ldots\times\llbracket 0, n_d-1\rrbracket \subseteq \ACN^t\}
\end{displaymath}
\end{definition}

When considering the specific case of the 2-dimensional von Neumann neighborhood, the real time is defined by $\tau_{\ACN_{\operatorname{vN}}}(n, m) = n + m - 2$. There is however a well known constant speed-up result :
\begin{proposition}[folklore]
	For any $k\in \NN$, any language that can be recognized in time $(\tau_{\ACN_{\operatorname{vN}}}+k)$ by a 2DCA working on the von Neumann neighborhood can also be recognized in real time by a 2DCA working on the von Neumann neighborhood.
\end{proposition}

So it will be enough to prove that a language is recognized in time $(n,m)\mapsto n+m+k$ for some constant $k$ to prove that it is recognized in real time.

\subsection{Polyominoes}

\begin{definition}[Polyomino]
	A \emph{placed polyomino} is a finite and 4-connected subset of $\ZZ^2$. A \emph{polyomino} is the equivalence class of a placed polyomino up to translation.
\end{definition}

\begin{definition}[HV-Convexity]
	A polyomino $p$ is said to be \emph{horizontally (resp. vertically) convex} if any cell between two cells of the polyomino on a same horizontal (resp. vertical) line is also a cell of the polyomino~:
	\begin{displaymath}
		\forall x_1, x_2, x_3, y \in \ZZ, \quad x_1\leq x_2\leq x_3 \wedge (x_1, y) \in p \wedge (x_3, y) \in p \Rightarrow (x_2, y)\in p
	\end{displaymath}
	
	A polyomino is \emph{HV-convex} if it is both horizontally and vertically convex (see Figure~\ref{fig:polyominoes}).
\end{definition}

\begin{figure}
	\centering
	\begin{minipage}{0.50\textwidth}
		\centering
		\includegraphics[scale=1,page=2]{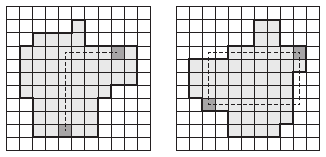}
		\caption{The polyomino on the left is L-convex (the figure shows an inner path connecting two cells with at most one direction change, and there is such a path for any pair of cells). The polyomino on the right is HV-convex but not L-convex as illustrated by the pair of highlighted cells for which there is no inner connecting path that changes direction at most once.}
		\label{fig:LPath}
	\end{minipage}\hfill
	\begin{minipage}{0.45\textwidth}
		\centering
		\includegraphics[scale=1, page=3]{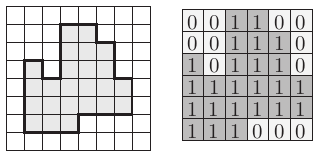}
		\caption{A polyomino (left) and its corresponding picture over $\{0,1\}$ (right). When this picture is encoded as a configuration of a cellular automaton, the origin of the automaton is on the lower left corner of the picture.}
		\label{fig:polyominoPicture}
	\end{minipage}
\end{figure}

We will now present the notion of L-convex polyomino, first introduced in \cite{castiglione03} to classify HV-convex polyominoes. Informally, an L-convex polyomino $p$ is such that for any two of its cells there exists a 4-connected path of cells of $p$ that connects them such that the path changes direction at most once (see Figure~\ref{fig:LPath}).

The following remarks will lead to a formal definition of L-convex polyominoes~:
\begin{itemize}
	\item a path that changes direction at most once connecting two cells of a polyomino $p$ on the same row (resp. column) is fully horizontal (resp. vertical) therefore L-convex polyominoes are HV-convex~;
	\item if $c_1=(x_1,y_1)$ and $c_2=(x_2,y_2)$ are two cells in an L-convex polyomino $p$, either $a_1 = (x_1,y_2)$ or $a_2=(x_2,y_1)$ is a cell of $p$ because $a_1$ and $a_2$ are the angles of the only two paths connecting $c_1$ and $c_2$ that change direction at most once~;
	\item if a polyomino $p$ is HV-convex and such that for any two of its cells $c_1=(x_1,y_1)$ and $c_2=(x_2,y_2)$ either $a_1=(x_1,y_2)$ or $a_2=(x_2,y_1)$ is a cell of $p$, then $p$ is L-convex since by HV-convexity, the whole path connecting $c_1$ to $c_2$ going through $a_1$ or $a_2$ is in $p$.
\end{itemize}

\begin{definition}[L-Convexity]
	A polyomino $p$ is L-convex if it is HV-convex and verifies
	\begin{displaymath}
		\forall x_1, x_2, y_1, y_2 \in \ZZ, \quad (x_1,y_1)\in p \wedge (x_2, y_2)\in p \Rightarrow (x_1, y_2) \in p \vee (x_2, y_1) \in p
	\end{displaymath}
\end{definition}

Given a polyomino $p$, the picture over the alphabet $\{0,1\}$ associated to $p$ is the picture whose dimensions are the dimensions of the minimal bounding rectangle of $p$, where the cell has state 1 if the corresponding cell is in the polyomino and 0 otherwise (see Figure~\ref{fig:polyominoPicture}). We define the language $L_{\operatorname{L-convex}}$ as the picture language of all L-convex polyomino pictures.

\section{Main Result}

This section will be entirely devoted to the proof of the following result
\begin{theorem}
	\label{theo:main}
	The picture language $L_{\operatorname{L-convex}}$ of L-convex polyomino pictures is recognizable in real time by a 2DCA working on the von Neumann neighborhood.
\end{theorem}

The proof will be done by describing the behavior of a 2DCA working on the von Neumann neighborhood that recognizes $L_{\operatorname{L-convex}}$ in real time. In this description we will use cardinal directions north, south, east and west to denote the different directions on the configuration as follows~:
\begin{itemize}
	\item north is towards the increasing $y$ axis~;
	\item south is towards the decreasing $y$ axis~;
	\item east is towards the increasing $x$ axis~;
	\item west is towards the decreasing $x$ axis.
\end{itemize}
With such conventions, the origin of the automaton is located at the south-west (SW) angle of the picture in the initial configuration and the picture therefore extends from the origin eastward and northward.

\subsection{Preliminary Check}

First of all, the automaton must check that the input is the picture of a HV-convex polyomino.

To do so, during the first step of the computation, each cell containing a $1$ considers its neighbors and remembers which of them also contains a $1$. Then a signal moves westward from the eastmost point of each row and southward from the northmost point on each column. These signals check that each row and each column contains exactly one segment of connected $1$ symbols. Moreover, the signals check that the segment of $1$ on each line and column is connected to that of the neighbor rows and columns using the neighboring information gathered during the first step.

These two properties guarantee that the polyomino is connected, HV-convex and that the picture's dimensions are that of the minimal bounding rectangle (no empty row or column). If an error is found on a row or column, the signal is directed towards the origin and the input is not accepted.

We can now assume that the input corresponds to a HV-convex polyomino picture, and must determine whether it is also L-convex.

\subsection{Characterization of L-Convex Polyominoes}

We will now present the characterization of L-convex polyominoes that will be used by the automaton. It is a slighly rephrased version of the characterization presented in \cite{brocchi13} (Theorem~2).

\begin{figure}[htb]
	\centering
	\includegraphics[scale=1, page=1]{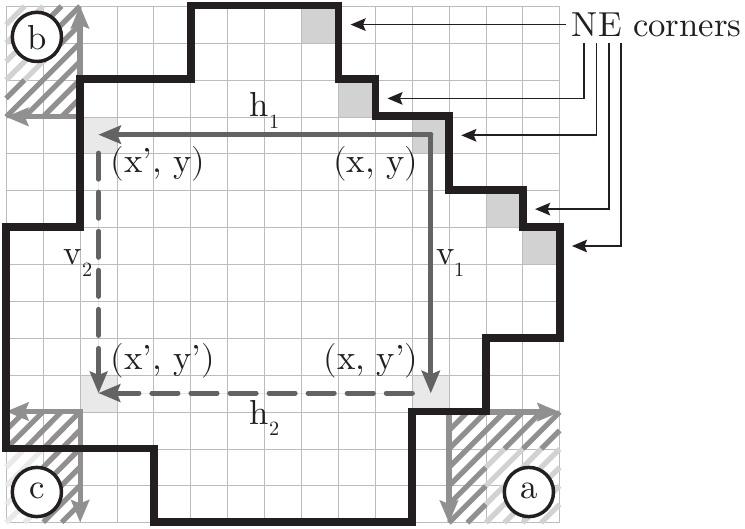}
	\caption{A HV-convex polyomino is L-convex is for any of its NE corners (represented as dark grey cells), no cell of the polyomino lies in any of the three zones represented in hatched light grey, and symmetrically for all of its NW corners. The illustrated polyomino is not L-convex because there are two cells in the lower left hatched area (these cells cannot be connected to the represented NE corner by an inner path with at most one direction change).}
	\label{fig:characterization}
\end{figure}

Given a polyomino $p$, we say that a cell of $p$ is a \emph{corner} if it has two consecutive neighbors that are not in $p$. We classify corners depending on the directions in which such neighbors not in $p$ are located~: a north-east (NE) corner is one such that the northern and eastern neighbors are not in $p$, and we similarly have NW , SW and SE corners (see Figure~\ref{fig:characterization} for an illustration of NE corners). Note that corner types are not exclusive~: a cell can for instance be both a NE and NW corner.

\begin{proposition}[Characterization of L-convex polyominoes \cite{brocchi13}]
	\label{prop:characterization}
	A HV-convex polyomino $p$ is L-convex if and only if for every NE corner $c=(x,y)$, denote by $(x, y')$ the southest cell of $p$ on the same column as $c$, and $(x', y)$ the westmost cell of $p$ on the same row as $c$, there is no cell $(x'', y'')$ of $p$ verifying any of the following three conditions
	\begin{enumerate}
		\renewcommand{\theenumi}{(\alph{enumi})}
		\item $x'' > x$ (resp. $x'' > x'$) and $y'' < y'$
		\item $x'' < x'$ (resp. $x'' < x'$) and $y'' > y'$
		\item $x'' < x'$ (resp. $x'' < x'$) and $y'' < y'$
	\end{enumerate}
	and the symmetric conditions holds for all NW corners (in the South and East directions).
\end{proposition}

Figure~\ref{fig:characterization} illustrates this characterization.

\begin{proof}[sketch]
	It is enough to verify that all pairs of corners of a HV-convex polyomino are connected by an inner path with at most one change of direction. Moreover by symmetry we can consider only NW and NE corners.
	
	The cells $(x',y)$, $(x,y')$ and $(x', y')$ in the characterization represent the farthest points that can be reached from a given corner in their respective directions. Cells of the three restricted areas cannot be connected to the corner and conversely all cells not in these areas can be connected to the corner.
\end{proof}

Note that because the polyomino is assumed to be HV-convex it is enough to check that there is no polyomino cell on the two lines extending from the starting check point (represented in dark hatched grey in Figure~\ref{fig:characterization}). For instance, for the condition $(a)$, it is enough to check that there is no cell $(x'', y'-1)$ with $x''>x$ and no cell $(x+1, y'')$ with $y'' < y'$ in the polyomino. This follows from the 4-connectedness of the polyomino.

Although the conditions to verify are perfectly symmetric for NE and NW corners, when implementing it on a real time cellular automaton the case of NE corners is significantly simpler because all signals move towards the origin at maximum speed so the result of the verification easily arrives on time. On the other hand, for NW corners, some signals move eastward (away from the origin) so it would take too much time to send the signal all the way to the east side and back to the origin. We will therefore now focus on implementing the characterization for NE corners and come back to the NW corners at the end of the proof.

\subsection{Compression and Marking}

The characterization from Proposition~\ref{prop:characterization} depends on cells being able to tell if there is a polyomino cell in a given direction from them. To make sure that each cell knows this information, consider signals going eastward from the west side of each row. If the initial configuration is the picture configuration of a HV-convex polyomino, there is exactly one segment of $1$ symbols on each row. Before the signal meets the first $1$, cells can be notified that there is no $1$ westward and that there is at least one $1$ eastward. On the segment of $1$, cells are notified of whether they are a border cell or an inner cell and, after the segment of $1$, all cells are notified that there is a $1$ westward and none eastward. Of course the same thing can be done on columns with northward signals.

\begin{figure}[htb]
	\centering
	\includegraphics[scale=1]{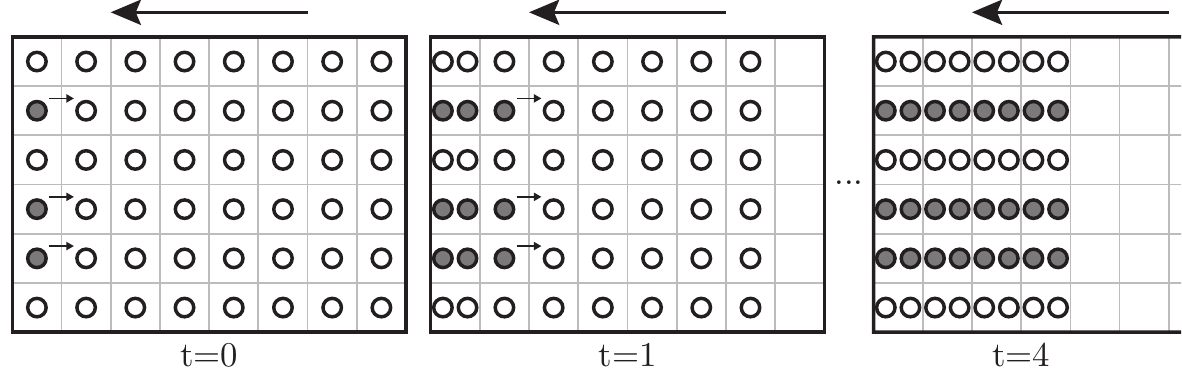}
	\caption{Horizontal compression of the input, with eastward transmission of information (dark dots).}
	\label{fig:compression}
\end{figure}

Now consider a horizontal compression of the input as illustrated by Figure~\ref{fig:compression}. To compress the input, consider that each cell can now hold two initial states instead of one (this can be done by increasing the number of states of the automaton) and move all states westward unless the column in which they should go is full (contains two states) or is out of the boudaries of the initial configuration (ignore the darker dots from Figure~\ref{fig:compression} for the moment).

Such a compression takes $\lceil\frac{n}{2}\rceil$ time steps where $n$ is the width of the input. During these steps, no signal can propagate westward because the initial data is already moving west at maximum speed but the time lost performing the compression can be recovered afterwards because each cell now sees twice as many states horizontally, which means that relatively to the original states, horizontal signals can perform two steps at a time.

During the compression, signals can however be propagated eastward (as illustrated by the darker dots in Figure~\ref{fig:compression}). This means that while the compression is taking place, the signal indicating to each cell if it has $1$ symbols east or west can propagate, so that at the end of the compression, cells have access to this information.

By performing a vertical compression after the horizontal one we can otain in half of the real time a compressed copy of the initial configuration on which every cell now has the added information of whether there is a $1$ in any of the four directions. Moreover, later in the construction we will need to know which columns correspond to the same horizontal segment on the southern border of the polyomino, so we also propagate northward signals  during the vertical compression from the borders of all horizontal segments of the southern border of the polyomino (see dashed northward arrows in Figure~\ref{fig:horizontalSegments}).

After both compressions, the computation of the automaton can properly start and in this computation horizontal and vertical signals can propagate twice as fast and all information is twice closer to the origin. This means that the compressed run of the automaton can behave exactly as if the configuration was not compressed but was given the extra information propagated by the eastward and northward signals from the beginning\footnote{This compression technique works in our case because the automaton (as it will be described later) only uses horizontal and vertical signals that change directions a bounded number of times. It is only possible to simulate two steps of the uncompressed automaton if they only involve horizontal or vertical movement, not both.}. We will now ignore the compression in the following explanations, and simply consider that the information propagated by the northward and eastward signals is readily available to each cell.

\textbf{Remark:} As it is described, it looks as if cells should know when the compression is finished to start performing the next task (be it the second compression or the accelerated simulation of the uncompressed automaton). However, one can show that cells can \emph{asynchronously} start the next task as soon as they have the necessary information to do so. It is sufficient to detect when all cells in their neighborhood have finished the compression to perform one step of the next task. From there, we can show that if each cell advances the following task as soon as it has enough information to do so, cells that have completed the compression early will be slowed down progressively to wait for the further cells to catch up. However, the last cells to finish the compression will never be slowed down as all other cells have the necessary information available to them. This means that by continuing the computation after the compression as soon as the information is available,  all cells are at least as advanced as if all had started their computation at the time when the compression is finished, thus negating the need to synchronize all cells after the compression.

\subsection{First Conditions of the Characterization}

With the informations we have, checking conditions $(a)$ and $(b)$ of Proposition~\ref{prop:characterization} is very easy as it is only a matter of sending a westward and a southward signal from each NE corner. When these signals reach the border of the polyomino, they check that there are no $1$ in the corresponding area by using the information that was transmitted to each cell during the compressions. If a $1$ is found where it should not be, a signal is directed towards the origin to indicate that the input should not be accepted.

There are no conflicting signals during this step because there can be at most one NE corner per column and one at most per row.

\subsection{The Third Condition}

The third condition from Proposition~\ref{prop:characterization} is much more complex to implement. It requires sending a westward signal $h_1$ and a southward signal $v_1$ from each NE corner and having these signals generate secondary signals $h_2$ (westward, from the collision of $v_1$ with the border of the polyomino) and and $v_2$ (southward from the collision of $h_1$ and the border). The intersection of $h_2$ and $v_2$ indicate the cell on which condition $(c)$ should be checked, as illustrated by Figure~\ref{fig:characterization}.

Two problems arise when implementing this behavior~:
\begin{itemize}
	\item although $v_1$ and $h_1$ signals originating from different NE corners will never overlap, if two signals arrive on the same row or column they will produce $v_2$ or $h_2$ signals that might overlap~;
	\item $h_2$ signals might intersect with many $v_2$ signals, but only one of them originates from the same NE corner. It is therefore necessary to ensure that the verification of condition $(c)$ is not performed on cells that do not correspond to a valid intersection of $h_2$ and $v_2$ signals.
\end{itemize}

\subsubsection{Priority Rule}

\begin{figure}[htb]
	\centering
	\includegraphics[scale=1]{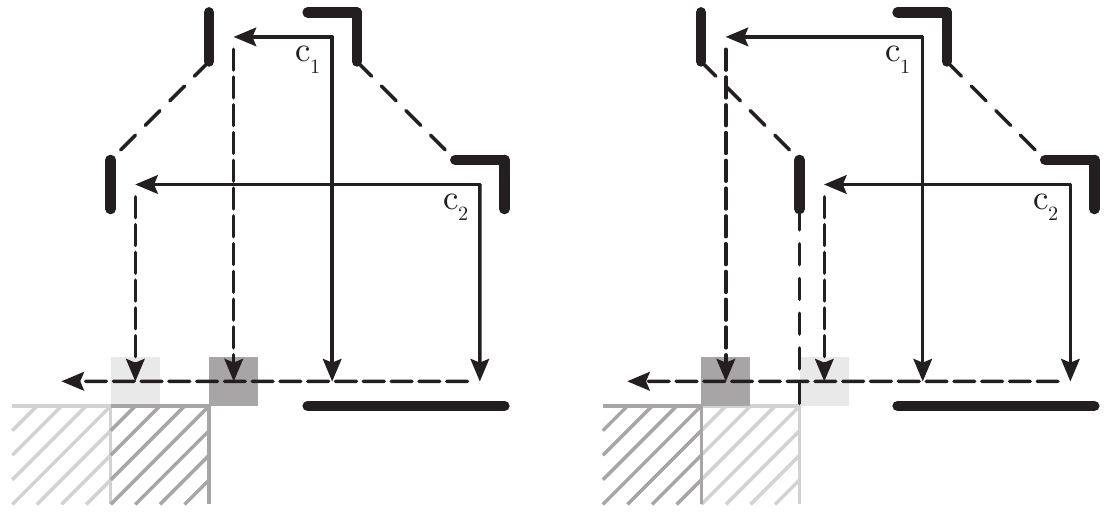}
	\caption{When two $v_1$ signals arrive on the same row we can always safely ignore the one originating from the eastmost NE corner.}
	\label{fig:horizontalConflict}
\end{figure}

To solve the first problem, we use a simple priority rule~: if two NE corners $c_1$ and $c_2$ are north of the same horizontal segment on the southern border of the polyomino, we can ignore the easternmost one. There are two cases to consider (illustrated by Figure~\ref{fig:horizontalConflict}). Assume $c_1$ lies north-west of $c_2$~:
\begin{itemize}
	\item if the westward $h_1$ signal from $c_1$ reaches the border east of that from $c_2$ (left of Figure~\ref{fig:horizontalConflict}), then the area that would be checked by considering the intersection of the signals $v_2$ and $h_2$ from $c_1$ (dark grey area in the Figure) is east of the one that would be considered by the intersection from $c_2$ (light grey area) and therefore contains it entirely, which means that it is not necessary to check the area indicated from $c_2$~;
	\item if on the contrary the $h_1$ signal from $c_1$ arrives west of that from $c_2$ (right of Figure~\ref{fig:horizontalConflict}), the HV-convexity of the polyomino guarantees that there can be no $1$ in either of the two areas considered by the intersections from $c_1$ and $c_2$ since there is at least one $1$ north west of where the $h_1$ signal from $c_2$ arrives, no $1$ west of that point so there cannot be any $1$ west and south of it. In this case, it doesn't matter which intersection is considered since neither will find a contradiction with the $(c)$ condition from Proposition~\ref{prop:characterization}.
\end{itemize}

A symmetrical argument shows that it is sufficient to consider signals originating from the southernmost of two NE corners whose $h_1$ signals arrive on the same vertical segment of the western border of the polyomino. Horizontal and vertical signals are however handled differently because the last part of the construction is not symmetrical.

\begin{figure}[htb]
	\centering
	\includegraphics[scale=1]{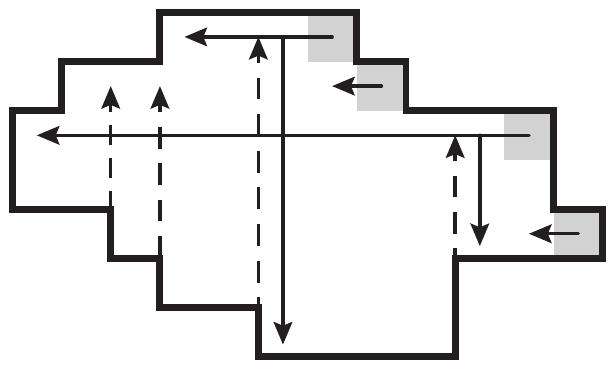}
	\caption{$h_1$ signals from an NE corner are interrupted if they detect that there is another NE corner west whose $v_1$ signal would arrive on the same horizontal segment on the southern border of the polyomino. $v_1$ signals are sent by $h_1$ signals on the westmost column corresponding to the southern horizontal segment.}
	\label{fig:horizontalSegments}
\end{figure}

We want to make sure that there are as many $v_1$ signals as there are distinct (non-overlapping) $h_2$ signals. To do so, $v_1$ signals are not sent directly by NE corners but rather sent by the $h_1$ signal when the $h_1$ signal knows that the corner it originated from is the westmost of the corresponding horizontal segment in the southern border (see Figure~\ref{fig:horizontalSegments}). When an $h_1$ signal finds a cell of the polyomino north before reaching the border of the southern segment (dashed line in the figure), it knows there is another NE corner west for that segment and therefore disappears. On the contrary, if such a signal reaches the border of the southern segment it sends the $v_1$ signal southward.

\subsubsection{Counters}

For $v_2$ signals, we need to solve the second problem that was described previously which is to determine which of the possibly many $h_2$ signals intersected is the one that originated from the same NE corner. To do so, $h_1$ signals produced by NE corners will count how many $v_2$ signals they cross while going west. If a $h_1$ signal crossed $n$ $v_2$ signals, then the $v_2$ signal it produces will consider that its corresponding $h_2$ signal is the $(n+1)$-th to last one (the last $n$ are not the one that should be considered).

\begin{figure}[htb]
	\centering
	\includegraphics[scale=1]{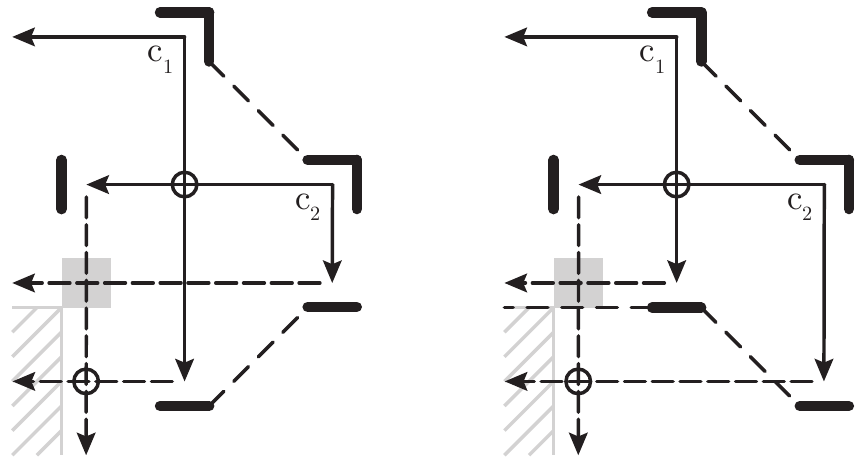}
	\caption{Considering that for each $v_2$ signal crossed by the $h_1$ signal from $c_2$, one of the last intersections with an $h_2$ signal should be ignored by the subsequent $v_2$ signal leads to a correct characterization.}
	\label{fig:counters}
\end{figure}

Figure~\ref{fig:counters} illustrates why the result of such a behavior is correct. Consider an NE corner $c_2$ such that its $h_1$ signal crossed the $v_1$ signal produced by an NE corner $c_1$ (top circled intersection)
\begin{itemize}
	\item if the $v_1$ signal produced by $c_2$ arrives north of the one produced by $c_1$ (left part of the figure) then the real intersection of the $v_2$ and $h_2$ signals from $c_2$ is the first that the $v_2$ signal from $c_2$ encounters, and the later one should be ignored (lower circled intersection)~;
	\item if on the contrary the $v_1$ signal from $c_2$ arrives south of that of $c_1$ (right part of the figure), the real intersection that should be considered is the last one but by considering the first the automaton will not find any contradiction to condition $(c)$ since by HV-convexity of the polyomino there are no $1$ south and west of either of the two intersections (so it will pick the wrong intersection but that will not change the final result).
\end{itemize}

In order to implement this rule, signals need to carry a binary counter. This counter should follow the signal at maximal speed, and will be incremented by the $h_1$ signal for each $v_1$ signal encountered (which can be easily done as incremental binary counters can be implemented on one-way one dimensional CA). For technical reasons, the counter has an initial value of 1.

As for the $v_2$ signal, as it crosses $h_2$ signals it checks if the area south-west of said intersection contains a $1$ (which can be done instantly because of the information gathered during the initial compressions), and if so decrements the counter\footnote{Decrementation can also be performed on a binary counter moving at maximal speed but in that case the length of the counter is not reduced when going from an $2^n$ to $(2^n-1)$, but instead a leading 0 is added, which is not a problem for our construction.}. If the counter is equal to 0 (decreasing a 0 counter leaves it at 0) when the $v_2$ signal reaches the southernmost border of the picture, no error is detected, but if it is positive then a message is sent to the origin to indicate that the input is not L-convex.

This works because we know that if there is no $1$ south-west of a cell $c$, there is no $1$ south-west of a cell south of $c$. If the $h_1$ signal crosses $n$ $v_1$ signals the counter indicates $(n+1)$ when the $v_2$ signal starts moving south and if the counter is at 0 it means that the $(n+1)$ last intersections were correct according to condition $(c)$ and therefore the $(n+1)$-th to last was correct.

Checking that the counter is 0 takes $\log(n)$ steps where $n$ is the maximal value of the counter ($\log(n)$ is the maximal length of the counter). If the counter is incremented to $n$ it means that there are at least $n$ NE corners north of the one from which the signal originated. This means that if the signal moves from this corner towards the origin (south or west) at maximal speed, it would reach the origin at least $n$ steps before the real time, and therefore it can spend $\log(n)$ steps checking the value of the counter and still arrive in real time.

Moreover, conflicts of overlapping counters can be resolved by the priority rule described previously. Precedence must always be given to the counter corresponding to the southernmost NE corner~:
\begin{itemize}
	\item when an $h_1$ signal reaches the west border of the polyomino, it marks the cell on which the $v_2$ signal is produced~;
	\item if a $v_2$ signal moves through such a marked cell, it is erased~;
	\item if the counter following a $v_2$ signal is on a cell where a new $v_2$ signal is created, the counter is invalidated (the end symbol is erased) so that the new $v_2$ signal has precedence over it. An invalidated counter will ignore decrementations and will ignore the test to 0 at the end.
\end{itemize}

\subsection{The North-West Corners}

The previous subsections describe how NE corners can properly implement the characterization of L-convex polyominoes from Proposition~\ref{prop:characterization}. NW corners will behave in a very similar fashion, but special care must be taken to prove that the result of their verification can reach the origin in real time.

On a regular configuration, a signal issued from a NW corner needs to go east through most of the polyomino, then south and then the result of the verification should travel back west to the origin. In doing so the signal goes twice through the width of the input which cannot be done in real time. To solve this problem, we consider the path of the signal during a horizontal compression of the configuration~:
\begin{itemize}
	\item the signal starts from the NW corner on the cell $c=(x, y)$~;
	\item during the $\frac x 2$ first steps the signal moves west with the compression, and when the cell is compressed, the $h_1$ signal is sent eastward~;
	\item meanwhile, the cell $(x',y)$ that should have been the target of the $h_1$ signal moves left with the compression. The $h_1$ signal and the cell arrive at the cell $(\frac{x'}{2},y)$ at time $(\frac{x'}{2},y)$~;
	\item the signal $v_2$ from $c$ moves south until it reaches the southern border of the input after $y$ steps. At this point a delay of at most $\log(h-y)$ steps is incurred to check the value of the counter ($h$ is the total height of the input)~;
	\item the result of the verification is directed towards the origin, it arrives at time $\frac{x'}{2} + y + \log(h-y)+ \frac{x'}{2} < x' + h$ which is before real time.
\end{itemize}

NW corners can therefore properly perform the necessary verifications to implement the characterization from Proposition~\ref{prop:characterization}, which concludes the proof of Theorem~\ref{theo:main}.

\bibliographystyle{plain}

\end{document}